\documentclass[aps]{revtex4-1}
\usepackage{amssymb}
\usepackage{amsmath}
\usepackage{graphicx}
\usepackage{epsfig}
\usepackage{subfigure}
\usepackage{amsfonts}

\begin{document}

\title{Experimental demonstration of quantum walks with initial superposition states}
\author{Qi-Ping Su$^{1}$, Yu Zhang$^{1}$, Li Yu$^{1}$, Jia-Qi Zhou$^{1}$, Jin-Shuang Jin$^{1}$,
Xiao-Qiang Xu$^{1}$, Shao-Jie Xiong$^{1,3}$, QingJun Xu$^{1}$, Zhe Sun$^{1}$, Kefei Chen$^{2}$,
Franco Nori$^{4,5 \ast}$}
\author{Chui-Ping Yang$^{1}$}
\email{Email: yangcp@hznu.edu.cn, Telephone number:  +86-571-28860503 office, +86-571-28865286  fax; and \\Email:
fnori@riken.jp, Telephone number:  +81-48-4679707 office, +81-48-4679681 secretary, +81-48-4679650 fax.}
\affiliation{$^{1}$Department of Physics, Hangzhou Normal University, Hangzhou, Zhejiang 310036, China}
\affiliation{$^{2}$Department of Mathematics, Hangzhou Normal University, Hangzhou 310036, China}
\affiliation{$^{3}$State Key Laboratory of Precision Spectroscopy, Department of Physics, East China Normal University, Shanghai 200062, China}
\affiliation{$^{4}$Theoretical Quantum Physics Laboratory, RIKEN Cluster for Pioneering Research,
Wako-shi, Saitama 351-0198,Japan}
\affiliation{$^{5}$Department of Physics, University of Michigan, Ann Arbor, Michigan 48109-1040,
USA}

\date{\today}

\begin{abstract}
The preparation of initial superposition states of discrete-time quantum walks (DTQWs) are necessary for the
study and applications of DTQWs. In linear optics, it is easy to
prepare initial superposition states of the coin, which are always encoded by polarization states; while the preparation of
superposition states of the walker is challenging.
Based on a novel encoding method,
we here propose a DTQW protocol in linear optics which enables the preparation of
arbitrary initial superposition states of the walker and the coin.
With this protocol, we report an
experimental demonstration of DTQW with the walker
initially in superposition states, by using only passive linear-optical elements.
The effects of the walker's different initial superposition states on the spread speed of the DTQW
and on the entanglement between the coin and the walker
are also experimentally investigated, which have not been reported before.
When the walker starts with superposition states, we show that the properties of DTQW are very different from
those of DTQW starting with a single position.
Our findings reveal different properties of DTQW and paves an avenue to study DTQW with arbitrary initial states.
Moreover, the encoding method enables one to encode an arbitrary
high-dimensional quantum state using a single physical qubit and may
be adopted to implement other quantum information tasks.
\end{abstract}
\maketitle
\date{\today}

\section{INTRODUCTION}

Quantum walks (QWs) are extensions of the
classical random walk \cite{s01} and have wide applications in
quantum algorithms \cite{s011,s02,s11}, quantum simulations
\cite{s03,s04,s041,s09,s0412,s0413}, quantum computation \cite{s042,s043,s05},
and so on \cite{s07,s12}. In standard one-dimensional (1D)
discrete-time quantum walks (DTQWs) \cite{s121,s102}, the walker's position can be
denoted as $|x\rangle$ ($x$ is an integer) and the coin can be
described with the basis states $|0\rangle_c$ and $|1\rangle_c$. The
evolutions of the walker and the coin are usually characterized by a
time-independent unitary operator $U=TS_c(\psi)$. In each step, the
coin is tossed by
\begin{equation}
S_c(\psi)=\left(
\begin{array}{cc}
\cos\psi & -\sin\psi  \\
\sin\psi  & \cos\psi%
\end{array}%
\right) \nonumber
\end{equation}
with $\psi\in(0^\circ,~90^\circ)$, and the walker is shifted by
$T=\sum_x|x+1\rangle\langle
x|\otimes|1\rangle_c\langle1|+\sum_x|x-1\rangle\langle
x|\otimes|0\rangle_c\langle0|$. In general, the result of the DTQW
with a finite number of steps is determined by the initial states of
the coin and the walker as well as the operator \textsl{U}. The
systematic investigation of DTQW properties and the effect of the initial
states on DTQW is important and necessary for applications of
DTQW.

It is worth noting that DTQW
with the walker initially in a superposition state has not been
demonstrated in experiments,
though there were several theoretical works \cite{s1021,s1022,s1023,s1024,s1025}.
DTQWs have been experimentally realized in several systems, such as
linear optics \cite{s08,s03,s11,s101,s105,s09}, ion traps
\cite{s15,s16}, cavity quantum electrodynamics \cite{s14}, and
neutral atom traps \cite{s17}.
In these DTQW experiments, the walker always starts from the original position
$|0\rangle$.
In most DTQW experiments, it is difficult to prepare an
arbitrary initial quantum superposition states of the walker.
Moreover, initial superposition states of the walker are
usually required for applications of DTQW. For example, an initial
uniform superposition state of the walker's position is required in
the Grover walk \cite{s011}. The DTQW with initial
superposition states of the walker are quite different from the standard DTQW
(i.e., the walker starts from a single position $|x\rangle$), because the final
state of the former is a coherent superposition of the final states of the standard DTQW
starting from different single positions and thus contains more
quantum interference.

Note that the effects of the coin's initial state on DTQW
have been well studied
\cite{s121,s06,s08,s101,s102,s105} but
the effects of the walker's initial superposition states on DTQW have not been investigated before in linear optics.
In previous linear-optical DTQW implementations
\cite{s171,s10,s18}, the coin was always
encoded in the 2-dimensional polarization space and the position of the walker
must be encoded in high-dimensional spaces.
In \cite{s18}, the coin
was encoded using the polarization state and the walker's positions
were encoded using the paths of the photons, in which an additional
calcite beam displacer was needed in each step, to perform the
conditional shift of the walker and DTQWs with up to 6 steps were measured in experiments.
In \cite{s10}, the coin was also
encoded using the polarization state and the walker's positions were
encoded using the temporal information of single photons, in which
the coin operator was adjustable and 5-step DTQWs were studied experimentally.
In linear optics, the preparation of arbitrary initial superposition states of the
coin is not difficult because it is easy to obtain arbitrary polarization states
using half-wave plates (HWPs) and quarter-wave plates (QWPs).
However, the preparation of arbitrary initial superposition states of the walker is
difficult.

To our knowledge, how to encode the walker's position using
the polarization state has not been reported in both theory and experiment.
It is difficult to encode a high-dimensional state using the
two-dimensional state.
If the walker's position is encoded using the polarization state,
arbitrary initial superposition states of the walker can be easily prepared in
linear-optical experiments.

In this work, we propose an encoding method to encode and read out high
dimensional states with a two-dimensional qubit.
With this method,
we report the first DTQW protocol with the walker's
position encoded in the polarization space of single photons. The DTQW protocol together with
the design of linear optical setup, in which the coin is encoded by two paths, allows us to arbitrarily set
the input state of the walker and coin.
Since the DTQW with initial superposition states of coin has been
well studied, here we focus on the experimental implementation of DTQW with the walker initially in superposition
states, which has not been reported before.
In this case, the effect of the walker's different
initial superposition states on the spread speed of the walker and on the
entanglement between the coin and the walker (i.e., the entropy of the
reduced density matrix of the coin) are studied
experimentally.

\section{RESULTS}
\textbf{DTQW Protocol in Linear Optics}

To implement DTQW in polarization space, we propose a novel encoding method in the Method section.
Here we make the specific choice of
encoding the walker's positions using non-orthogonal states
$|k\rangle_p=\cos\theta_k|H\rangle+e^{i\phi_k}\sin\theta_k|V\rangle$
(with integer $k$) in the polarization space of single
photons, by setting $\theta_k=k\Delta\theta$ and $\phi_k=0$.
The coin is encoded by two
paths $|0\rangle$ and $|1\rangle$ of the single photons. As shown in
Fig.~1(a), the operator for the conditional translation
of $|k\rangle_p$ should be expressed as
$$T_p=\sum_k|k+1\rangle_p\langle
k|\otimes|1\rangle\langle 1|+\sum_k|k-1\rangle_p\langle
k|\otimes|0\rangle\langle 0|.$$
In this case, the coin operator
$S_c$ can be implemented by a $\cos^2\psi/\sin^2\psi$ beam splitter
(BS) and the operator $T_p$ can be implemented via two HWPs oriented at $0^o$ and two HWPs oriented at
$\Delta\theta/2$ (step forward) and $-\Delta\theta/2$ (step back),
as shown in Fig.~1(b). Note that each HWPs oriented at
$0^o$ can be replaced by a mirror.

In each step of DTQW in polarization space, the state of the
system can be expressed as
$|\Psi_0\rangle_p|0\rangle+|\Psi_1\rangle_p|1\rangle$, with
\begin{equation}
\begin{aligned}
|\Psi_0\rangle_p=\frac{1}{N_p}\sum_{k=-n}^n a_k
|k\rangle_p=\frac{1}{N_a}\sum_{k=-n}^n a'_k
|k\rangle_p,\\
|\Psi_1\rangle_p=\frac{1}{N_p}\sum_{k=-n}^n b_k
|k\rangle_p=\frac{1}{N_b}\sum_{k=-n}^n b'_k |k\rangle_p,
\end{aligned}
\end{equation}
where $a'_k=a_k/C_a$, $b'_k=b_k/C_b$, $C_a=\sqrt{\sum|a_k|^2}$,
$C_b=\sqrt{\sum|b_k|^2}$, $N_a=N_p/C_a$, $N_b=N_p/C_b$, and $N_p$ is
the normalization coefficient, which guarantees
$\sum_k|a_k|^2+\sum_k|b_k|^2=1$. If one derives all values of $a_k$ and
$b_k$, the probability distribution of the DTQW can be obtained as
$|a_k|^2+|b_k^2|$.

The values of $a_k$ ($b_k$) can be obtained by using $a_k=C_a a'_k$
($b_k=C_b b'_k$). Note that $|\Psi_0\rangle_p$ ($|\Psi_1\rangle_p$)
has the same form as that of $|\Psi\rangle_q$ in the Eq. (\ref{e2}) and
the condition $\sum_k|a'_k|^2=1$ ($\sum_k|b'_k|^2=1$) is satisfied.
By using the encoding method introduced in the section II, $a'_k$
and $b'_k$ can be obtained one at a time by implementing the DTQW with different
angles $\Delta\theta$ and measuring $|\Psi_0\rangle_p$
and $|\Psi_1\rangle_p$ in path $|0\rangle$ and path $|1\rangle$,
respectively. Next, $C_a$ and $C_b$ can be obtained by solving two
equations
\begin{equation}
C^2_a+C^2_b=1~~~ \text{and}~~~ \frac{C^2_a}{C^2_b}\cdot\frac{|\sum_k a'_k\cos{(k\theta)}|^2}{|\sum_k
b'_k\sin{(k\theta)}|^2}=r,
\end{equation}
where $r=|\sum_k
a_k\cos{(k\theta)}|^2/|\sum_k b_k\sin{(k\theta)}|^2$ is the ratio
between the number of photons counted in path $|0\rangle$ and that
counted in path $|1\rangle$ at the same time period for an arbitrary
$\theta\neq0$. With this protocol, DTQW with the walker initially in an arbitrary superposition
state can be easily implemented.
\\

\textbf{Experimental demonstrations}

A feasible linear-optical setup for DTQW
in the photon polarization space is shown in Fig.~2. A
pair of photons are generated by the type-I spontaneous parametric down
conversion in a 3-mm-thick nonlinear beta-barium borate (BBO)
crystal pumped by a 100 mW diode laser (centered at 405.8 nm). One
photon is directly detected by detector $D_0$ as the trigger and the
other photon is adopted to make DTQW. Then an
arbitrary coin state can be initialized by a combination of one
polarizing beam splitter (PBS), two quarter-wave plates (QWPs), and
one HWP. An arbitrary state of the walker's position is represented as a
polarization state, which can be easily initialized by using two
HWPs on the two pathes of the DTQW photon after the PBS.

In Fig.~3, we show the results of the 1D Hadamard (i.e.,
setting $\psi=\pi/4$) DTQW with an initial state of
$(0.8|-1\rangle_p+0.6|1\rangle_p)|0\rangle$. Besides the probability
distribution of the walker's position $|a_k|^2+|b_k|^2$, the
coefficients $a_k$ and $b_k$ of the states $|\Psi_0\rangle_p$ and
$|\Psi_1\rangle_p$ for a 2-step DTQW, a 4-step DTQW, and a 6-step
DTQW are obtained and compared with the theoretical results. Note
that the error bars only indicate statistical uncertainty, which are
obtained by numerical simulations. Our results indicate that the
protocol introduced here is realizable in experiments and DTQW with
small errors can be implemented by using the setup shown in
Fig.~2.

Since the walker is initially in a superposition state of positions
$|-1\rangle_p$ and $|1\rangle_p$, the outermost positions of the
walker after an $n$-step DTQW are denoted by $|-n-1\rangle_p$ and
$|n+1\rangle_p$. To simplify the calculation of $a_k$ and $b_k$, we
have used $a_{n+1}=0$ and $b_{-n-1}=0$ for this type of
DTQW. In this case, only $n$ different $\Delta\theta$ are needed to
obtain $a_k$ and $b_k$ for an $n$-step DTQW.

The initial state of the walker's position can be easily prepared by
rotating the angles of the HWPs in our experiment. We have also
experimentally studied the DTQW with the initial state
$(0.6|-2\rangle_p+|0\rangle_p+0.8|2\rangle_p)|0\rangle/\sqrt{2}$.
The results of the reconstructed $a_k$, $b_k$ and $|a_k|^2+|b_k|^2$
after 6-step DTQW are shown in Fig.~4, which also fit well
with the theoretical results.

By means of this DTQW experiment, we also
study the relation between the spread speed $s$ and the initial
state of the walker's position. The spread speed $s$ is defined as
$s(n)=[\sigma(n)-\sigma(0)]/n$, where $\sigma(n)$ is the variance of
the position for an $n$-step DTQW. The results for a 4-step DTQW are
obtained and plotted in Fig.~5(a), in which the initial state is
set as
$(\alpha|-1\rangle_p+\sqrt{1-\alpha^2}|1\rangle_p)(|0\rangle+i|1\rangle)/\sqrt{2}$,
with $\alpha\in(-1,1)$. It is shown that different initial states of
the walker's position will lead to different spread speeds of DTQW.
Especially, for the DTQWs with $\alpha$ and $-\alpha$
($\alpha\neq0$), they have different $\sigma(n)$ and $s(n)$ ($n>0$)
though starting with the same $\sigma(0)$, which is quite different
from the classical random walk as shown in Fig.~5(a), because of
the quantum interference.

In Fig.~5(a), we also give the simulated relation of $s\sim\alpha$ for a 16-step and
50-step DTQW. It is shown that the properties of the
DTQW starting from superposition states of walker are very different
from that starting from a single position of walker. For example,
the parameter $s$ for $\alpha\neq0$ approaches a constant more slowly than that for
$\alpha=0$ as the number of DTQW steps increases.

A significant advantage of our DTQW protocol is that the whole final state
(and the density matrix) of the coin and the walker's position can
be obtained, rather than the position distributions only in other
DTQW experimental schemes. In this case, we can investigate the
entanglement between the coin and the walker in experiments by
calculating the entropy of the reduced density matrix of the coin.
The entropy is defined as $E=-\text{Tr}(\rho_c\ln\rho_c)$, where
$\rho_c=\text{Tr}_x(\rho_{cx})$ is obtained by tracing out the walker's position. In Fig.~5(b), a relation between
the entropy $E$ and the walker's initial state from the experiment
is compared with the theoretical results for 4-step DTQW. The
initial states used here are the same as that for Fig.~5(a),
whose entropies are $0$. We show that different initial states
of the walker will lead to different entropies after DTQW steps. The
results for 16-step and 50-step DTQW are also plotted in
Fig.~5(b). As the number of DTQW steps increases, the entropy
will approach a constant value rapidly for all $\alpha$, not only for
$\alpha=0$ \cite{s27}. Comparing Fig.~5(a) with Fig.~5(b),
it is clear that in general a larger spread speed of DTQW corresponds to a
larger entanglement between the coin and the walker.

\section{Discussion}
To implement DTQW for larger
steps, the errors of the experiment should be analyzed and improved.
The errors are related to the following factors:

(i) \textbf{The number of coincidence counts.} Because of the leakage of
photons at BS 1, the number of coincidence counts will decrease
rapidly with increasing the DTQW steps. The ratio of reflection to
transmission of all beam splitters adopted in our experiment is
50:50. The loss of detected photons after every two steps of DTQW
is about $70\%$. It is an advantage of two optical loops adopted in
the setup: photons are leaked and detected after every two steps
rather than after each step.

The errors in the experiment can be reduced by increasing the
number of coincident counts. One method is to replace the 50:50
coupler (i.e., BS 1) by a coupler with better ratios. The other
method is to increase the number of input photons. This can be
achieved in several ways. For example, the BBO can be replaced by a
high-efficiency periodically poled KTiOPO4 (PPKTP), the power of
the laser can be improved, or weak laser pulses can be directly used
 as the input photons of DTQW.

(ii) \textbf{The quality of the interference in the optical loops.} Quantum
walks are different from classical random walks because of
quantum interference. There are two optical loops in the setup. To
guarantee good interference of DTQW, optical paths of anticlockwise
and clockwise must be adjusted to be the same in each loop and two
loops must be adjusted to match each other. In general, the
quality of the interference will become worse when increasing walk steps.
An important factor leading to interference degradation is
the imperfections of devices, such as the reflection-dependence of
the polarization and nonplanar optical surfaces \cite{s18,s10}.

To show the scalability of this DTQW protocol and the setup,
we estimate the maximum number of steps that are in principle
achievable. We assume the use of perfect devices
(including detectors). In addition, we assume that the loss of detected photons after
every two steps of DTQW is $70\%$. By replacing the 50:50 coupler
with a 99:1 coupler and replacing the input single photons with laser
pulses of $1.4$ W power (5 MHz), there should be $\sim10,000$
photons detected in one second after 150 steps, provided that the
signal-to-noise ratio can be improved by adding an active switch to
couple the photons out of the loops.

In summary, we have experimentally demonstrated a 1D DTQW with the
walker moving in the polarization space of single photons. This
work differs from previous works in that the walker is initially in
a superposition state and the walker's position is encoded
using the photon polarization state.
The ability to operate with arbitrary initial superposition states
of DTQW and to implement DTQWs in a low-dimensional space opens an avenue
for not only discovering more properties of DTQWs but also more DTQW applications.
This DTQW protocol is quite general and can be
applied to other quantum systems (e.g., cavity or circuit QED
systems). This work may be extended to realize
multi-dimensional DTQW, which is of importance for
large-scale quantum computing based on quantum walks.
Moreover, the proposed encoding method enables one to encode an arbitrary
high-dimensional quantum state using a single physical qubit and may
be adopted to implement other quantum information tasks.

\section{METHODS}

\textbf{The encoding method}

Suppose there is an $n$-dimensional state
\begin{equation}
|\Psi\rangle=\sum_{k=1}^{n}a_k|k\rangle~~ \textrm{with}~~
~\sum_{k=1}^{n}|a_k|^2=1,\label{e1}
\end{equation}
in which $|k\rangle$ ($k=1,2,...,n$) are orthogonal states. In the
following, we propose a method to encode this state with a
two-dimensional qubit and to read out all $a_k$ from the qubit by
measurement. The obtained $a_k$ can be used to reconstruct the
$n$-dimensional state $|\Psi\rangle$ of Eq. (\ref{e1}).

The state $|\Psi\rangle$ can be encoded by using $n$ non-orthogonal
states
$|k\rangle_q=\cos\theta_k|0\rangle+e^{i\phi_k}\sin\theta_k|1\rangle$
($k=1,2,...,n$) of a qubit, which correspond to $n$ different points
in the Bloch sphere of the qubit. With $|k\rangle$ replaced by
$|k\rangle_q$, the state of the qubit in the encoded state can be
expressed as
\begin{equation}
\begin{aligned}
|\Psi\rangle_q=\frac{1}{N_q}\sum_{k=1}^{n}a_k|k\rangle_q=\frac{1}{N_q}(C_{0}|0\rangle+C_{1}|1\rangle)~,
\label{e2}
\end{aligned}
\end{equation}
where $C_{0}=\sum_{k=1}^{n}a_k\cos\theta_k$,
$C_{1}=\sum_{k=1}^{n}a_k e^{i\phi_k}\sin\theta_k$, and
$N_q=\sqrt{|C_{0}|^2+|C_{1}|^2}$. Note that the same state
$|\Psi\rangle_q$ can be constructed by performing suitable
operations on an initial state of the qubit, which is just the case
considered in the DTQW protocol below.

We now show how to read out $a_k$ from $|\Psi\rangle_q$. The density
matrix of the encoded state $|\Psi\rangle_q$ can be expressed as
\begin{equation}
\rho_q=\frac{1}{N_q^2}\left(
\begin{array}{cc}
C_{0}^*C_{0} & C_{1}^*C_{0}  \\
C_{0}^*C_{1}  & C_{1}^*C_{1}%
\end{array}%
\right) ,
\end{equation}
which can be easily obtained by using tomographic measurement on the
qubit. Then one homogeneous equation of $a_k$ can be obtained as
\begin{equation}
\begin{aligned}
C_0-R\cdot C_1=0 \Rightarrow \sum_{k=1}^{n}\left(\cos\theta_k-R\cdot
e^{i\phi_k}\sin\theta_k\right)\cdot a_k=0,
\end{aligned}
\end{equation}
where $R=C_{0}/C_{1}$ is obtained from the measured $\rho_q$ as
\begin{equation}
R=\rho_{q11}/\rho_{q21}=\rho_{q12}/\rho_{q22}.\label{R}
\end{equation}

In this way, $n-1$ homogeneous equations for $a_k$ can be obtained
by repeatedly encoding the state $|\Psi\rangle$ onto a qubit with
$n-1$ different sets
$\{|1\rangle_q^j,|2\rangle_q^j,...,|n\rangle_q^j\}$ (where
$j=1,2,...,n-1$, and
$|k\rangle_q^j=\cos\theta_k^j|0\rangle+e^{i\phi_k^j}\sin\theta_k^j|1\rangle$),
and by measuring the corresponding $R^j$ one at a time. Then $a_k$
($k=1,...,n$) can be solved from the $n-1$ homogeneous equations
\begin{equation}
\sum_{k=1}^{n}\left(\cos\theta_k^j-R^j\cdot
e^{i\phi_k^j}\sin\theta_k^j\right)\cdot a_k=0\label{Rs}
\end{equation}
and the normalization condition $\sum_{k=1}^{n}|a_k|^2=1$. Now the
$n$-dimensional encoded state is reconstructed.

To exhibit the generality and scalability of this encoding method,
we calculated the fidelity between random 50-dimensional encoded
states and their reconstructed states from numerical simulations of the encoding
process, in which random errors of measured $R^j$
were assumed to be within $\pm10\%$ and a specific choice of the non-orthogonal states
\begin{equation}
\begin{aligned}
|k\rangle_q^j=\cos(jk\Delta\theta)|0\rangle+e^{ik\Delta\phi}\sin(jk\Delta\theta)|1\rangle
\end{aligned}\label{state}
\end{equation}
was used (i.e., $\theta_k^j=jk\Delta\theta$ and
$\phi_k^j=k\Delta\phi$). The averaged fidelity over 1000 random
simulations is obtained and its dependence on $\Delta\theta$ and
$\Delta\phi$ is plotted in Fig.~6(a). The averaged fidelity is larger than $90\%$
in the $\sim41\%$ area of
$\Delta\theta\in(0,180^o)$ and $\Delta\phi\in(0,360^o)$. It is
not hard to find suitable $\theta_k$ and $\phi_k$ to implement the
encoding method, and high fidelity can be achieved for encoding
large dimensional states if measurement errors are not very large.

Moreover, we experimentally demonstrated the encoding and reading of a
16-dimensional state in linear optics system. At first, we encode a
uniform state $\Phi_{16}=\frac{1}{4}\sum_{k=1}^{16}|k\rangle$ in the
2-dimensional polarization space of single photons by using
$\Delta\theta=22^o$ and $\Delta\phi=12^o$ in Eq.~(\ref{state}). This encoding can be
easily implemented by injecting photons into an adjusted setup
consisting of two QWPs and one HWP, which can be used to construct
an arbitrary one-qubit gate in the polarization space. Then density matrixes of the encoding
photons can be measured and the initial uniform state can be
reconstructed. The experimental result of the reconstructed state
is shown in Fig.~6(b), and a high fidelity of $99.1\%$ is
achieved.

The encoding method introduced here is feasible
for the present technology of linear optics and can be easily
applied in other quantum systems. One key advantage for the use of
this encoding method is the simplicity of setup, because only one
physical qubit is needed.
\\

\section*{Acknowledgments} This work is supported in part by the
NKRDP of China (Grant No. 2016YFA0301802) and the National Natural
Science Foundation of China under Grant Nos [11504075, 11374083,
11774076, 11375003, 11775065]. This work is partially supported by the MURI
Center for Dynamic Magneto-Optics via the AFOSR
Award No. FA9550-14-1-0040, the Army Research Office
(ARO) under grant number 73315PH, the AOARD grant
No. FA2386-18-1-4045, the IMPACT program of JST,
CREST Grant No. JPMJCR1676, JSPS-RFBR Grant
No. 17-52-50023.

\section*{Conflict of interests}
The authors declare that they have no conflict of interest.

\section*{contributions}
Q.P.Su devised various aspects of the project and designed the
experimental methodology. Q.P.Su, Y.Zhang, J.Q.Zhou and S.J.Xiong carried out
the experiment and analyzed the data.
Q.P.Su, L.Yu, C.P.Yang, J.S.Jin and X.Q.Xu developed the theoretical aspects.
Q.P.Su, C.P.Yang and F.Nori wrote the
manuscript, with contributions from K.F.Chen, Z.Sun and Q.J.Xu.
All authors discussed the results and contributed to refining the
manuscript.

\newpage
\section*{Figure legends}

\subsection*{Fig. 1}
(a) Encoding of the walker's position and
the implementation of conditional translation $T_p$ in the
polarization space. $|H\rangle$ ($|V\rangle$) is the horizontal
(vertical) polarization state. (b) Optical realization of the
time-independent evolution operator $U=T_pS_c$ for DTQW, in which
the coin is tossed by the beam splitter (BS) and the conditional
transition of the walker's positions $|k\rangle_p$ is implemented by two
half-wave plates (HWPs) oriented at $\Delta\theta/2$ (for the coin
state $|0\rangle$) and $-\Delta\theta/2$ (for the coin state
$|1\rangle$) and two $0^o$ HWPs.

\subsection*{Fig. 2}
Schematic of the experimental setup for
quantum walk in the polarization space, in which $D_0$ is a
single-photon detector and $D_1$ ($D_2$) represents the standard
polarization analysis detector. A pair of photons are generated in
the BBO crystal. One photon is detected directly by $D_0$ as the
trigger and another photon is adopted to make the DTQW. The initial
states of the coin and the walker are prepared in the light yellow block.
The DTQW is realized with two optical
loops. The photons experience different steps of DTQW and can be
distinguished by their arrival times at $D_1$ or $D_2$ after a
trigger event. PBS: polarizing beam splitter.

\subsection*{Fig. 3}
Theoretical and experimental results of the
1D Hadamard DTQW after 2, 4, and 6 steps with an initial state of
$(0.8|-1\rangle_p+0.6|1\rangle_p)|0\rangle$. The green (blue) bars
represent the experimental (theoretical) results. The red error bars
represent statistical errors.

\subsection*{Fig. 4}
Theoretical and experimental results for the
6-step DTQW with an initial state of
$(0.6|-2\rangle_p+|0\rangle_p+0.8|2\rangle_p)|0\rangle/\sqrt{2}$.
The green (blue) bars show the experimental (theoretical)
results. The red error bars represent statistical errors.

\subsection*{Fig. 5}
(a) Relation between the spread speed $s$
and the initial state of the walker's position for 4-step (red),
16-step (blue) and 50-step (green) DTQW.
The thin solid curves are from the corresponding classical random walks for 4, 16, and 50 steps, which
have a symmetry on the two sides of $\alpha=0$. (b) Relation between the entropy $E$ and the
initial state of the walker's position for 4-step (red), 16-step (blue)
and 50-step (green) DTQW. The initial state of the DTQW is assumed as
$(\alpha|-1\rangle_p+\sqrt{1-\alpha^2}|1\rangle_p)(|0\rangle+i|1\rangle)/\sqrt{2}$.
The curves are from our theoretical calculations.
The red dots show the experimental data and the red error bars
represent statistical errors.

\subsection*{Fig. 6}
(a) Numerical simulations for the fidelity of
the encoding of 50-dimensional states into qubits. (b) Experimental results for the reconstruction
of a 16-dimensional state $\frac{1}{4}\sum_{k=1}^{16}|k\rangle$ from
encoded single photons in a linear-optical system.
\\

\newpage
\section*{Figures}

\begin{figure}[h]
\begin{center}
\includegraphics[width=14 cm, clip]{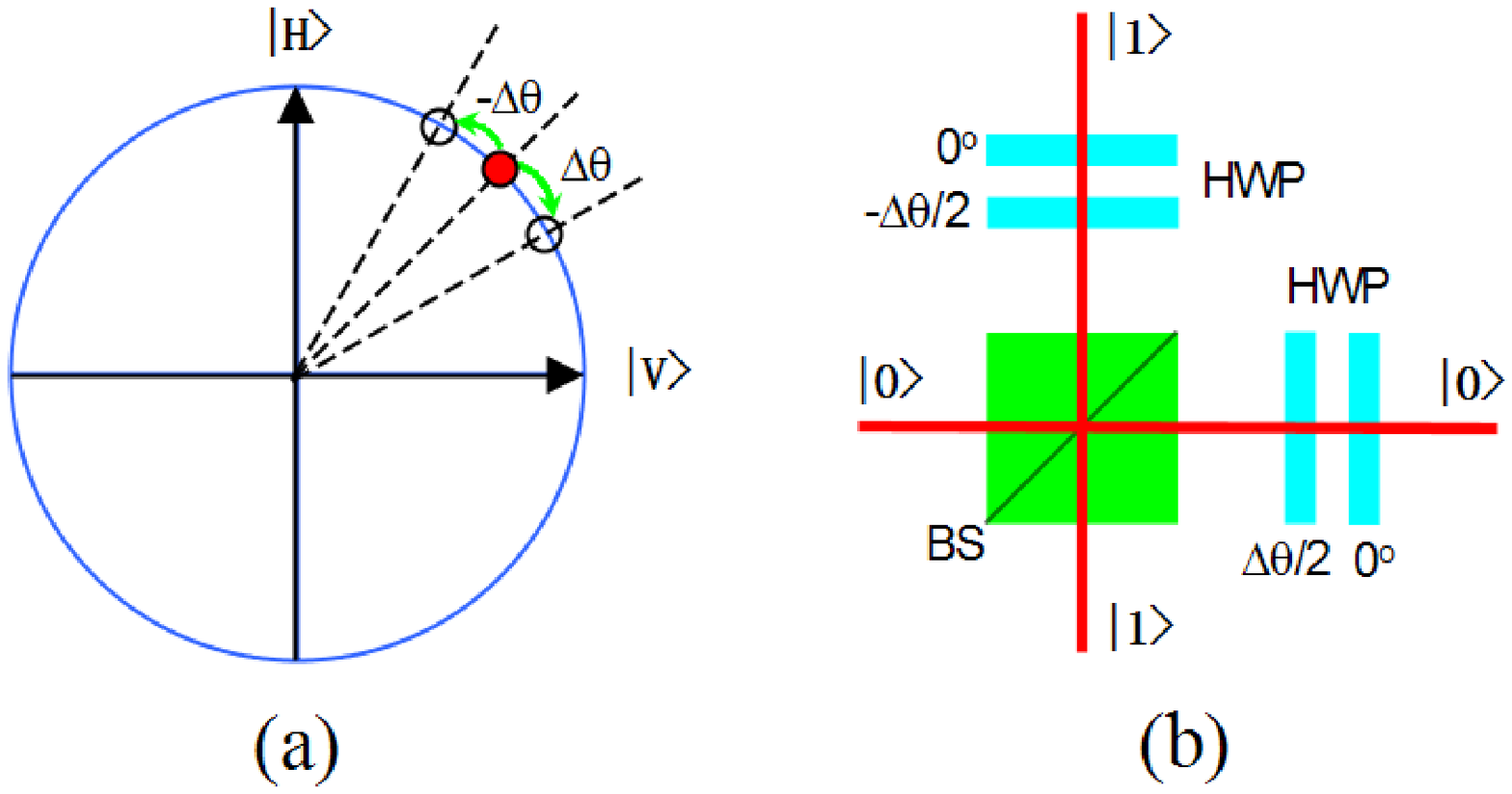}
\caption{}
\end{center}
\end{figure}

\begin{figure}[b]
\begin{center}
\includegraphics[width=14 cm, clip]{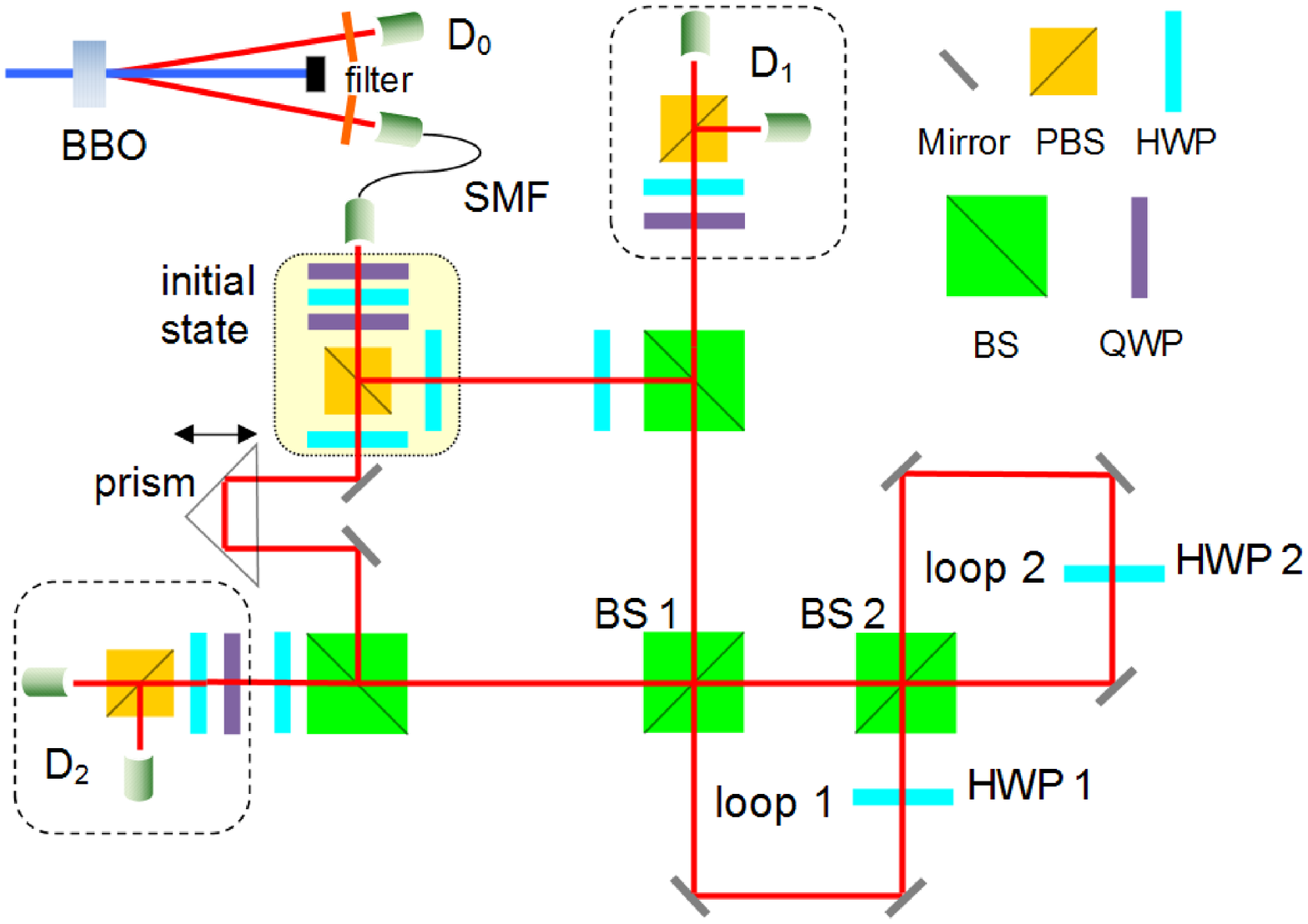}
\caption{}
\end{center}
\end{figure}

\begin{figure}[t]
\begin{center}
\includegraphics[width=17 cm, clip]{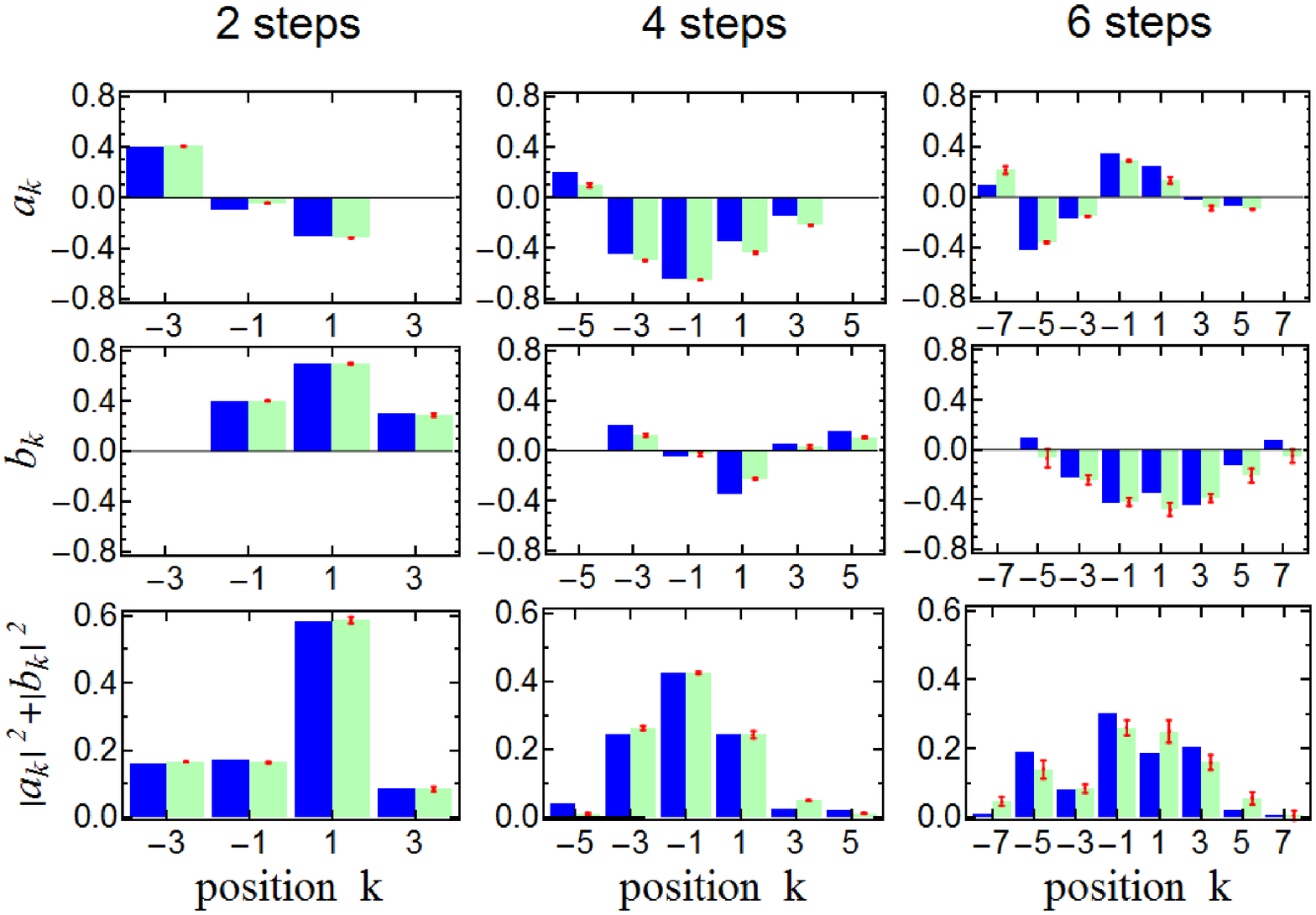}
\caption{}
\end{center}
\end{figure}

\begin{figure}[t]
\begin{center}
\includegraphics[width=17 cm, clip]{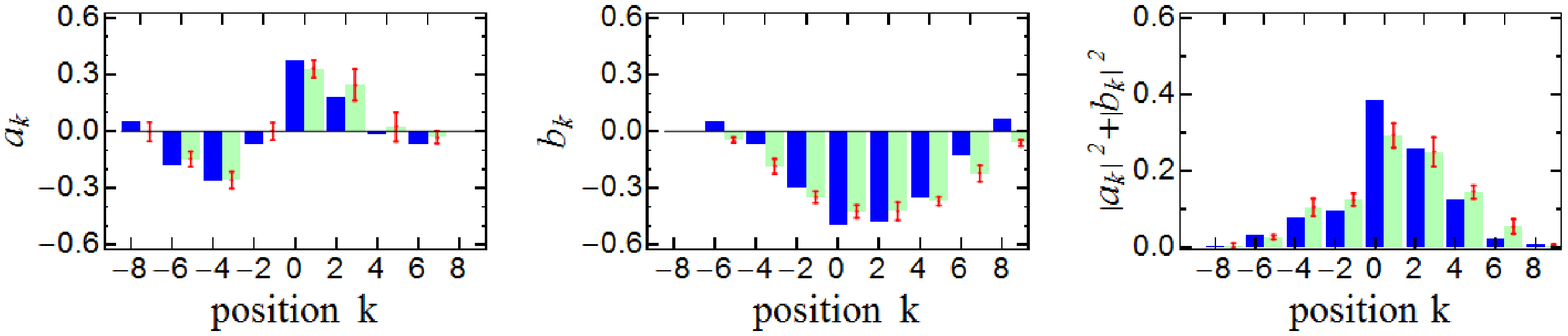}
\caption{}
\end{center}
\end{figure}

\begin{figure}[t]
\begin{center}
\includegraphics[width=17 cm, clip]{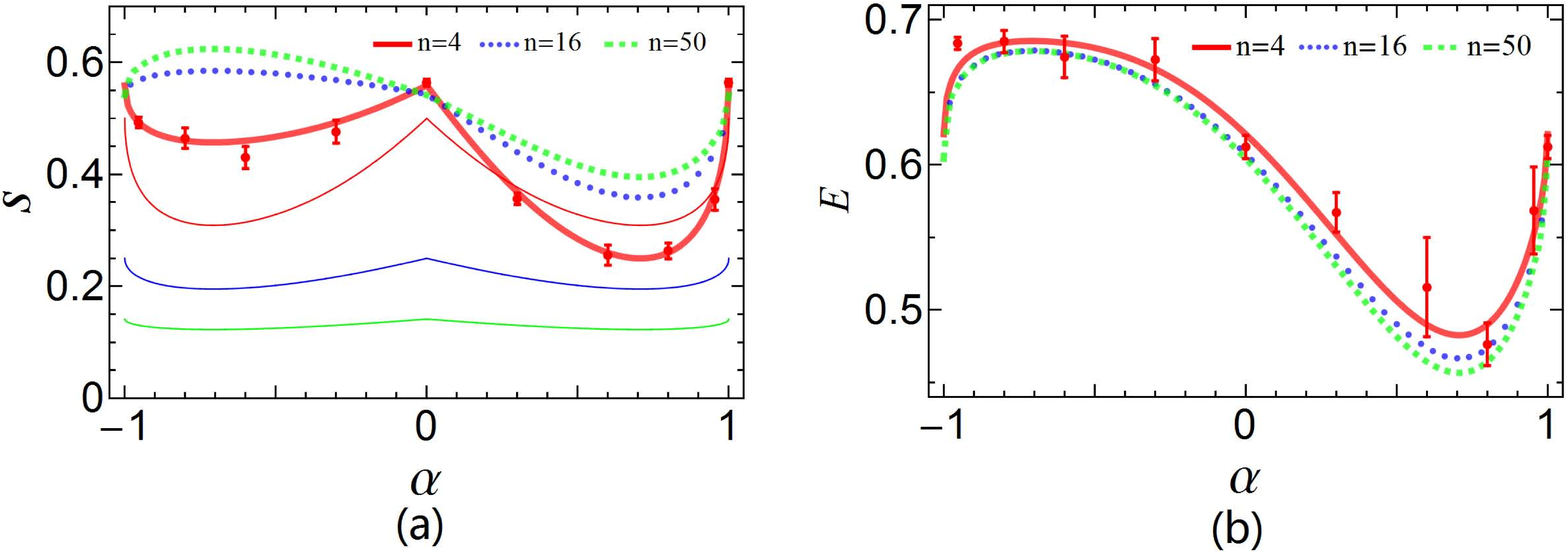}
\caption{}
\end{center}
\end{figure}

\begin{figure}[t]
\begin{center}
\includegraphics[width=17 cm, clip]{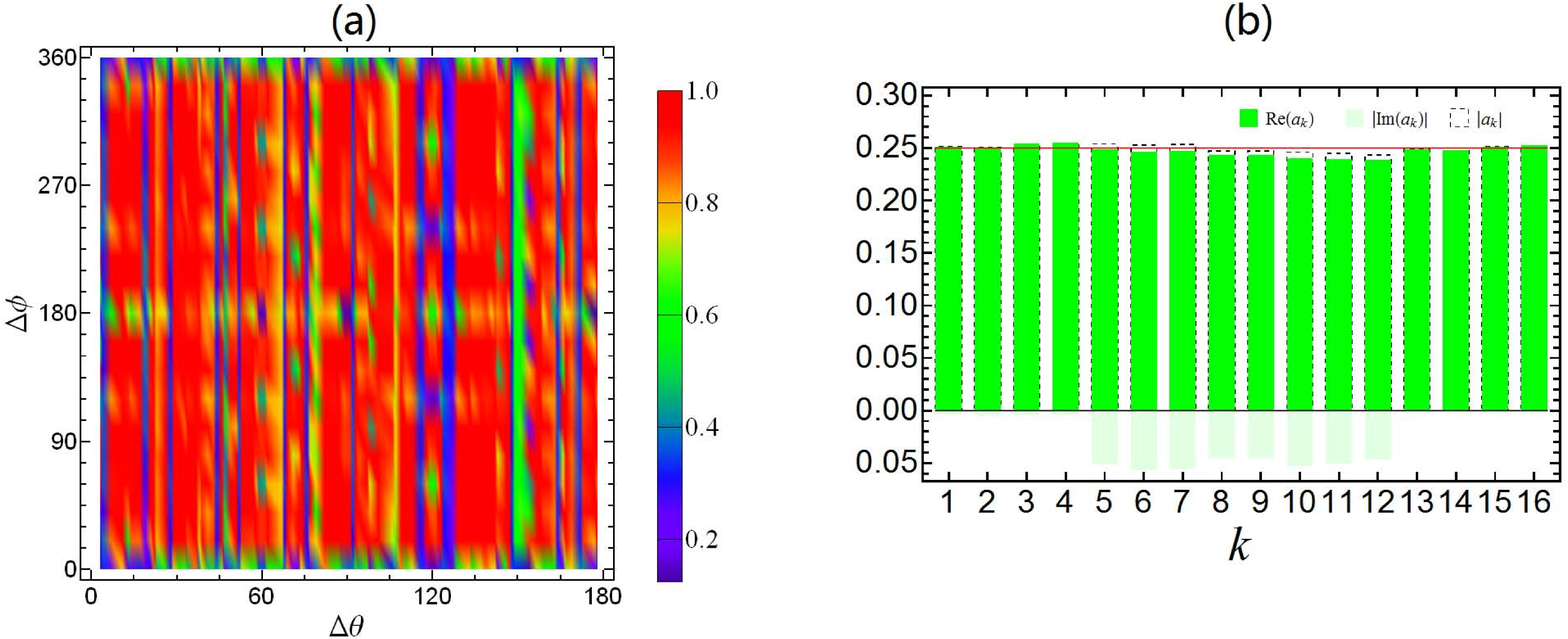}
\caption{}
\end{center}
\end{figure}

\end{document}